\documentclass{jps-cp}

\usepackage{graphicx}

\newcommand{\La}{{\Lambda}}
\newcommand{\Si}{{\Sigma}}

\title{Hyperon single-particle potentials in nuclear matter based on baryon-baryon
interactions derived within chiral effective field theory}

\author{Asanosuke \textsc{Jinno}$^{1}$,
Johann \textsc{Haidenbauer}$^{2}$,
and Ulf-G. \textsc{Mei}{\ss}\textsc{ner}$^{3,2}$
}

\inst{$^{1}$Department of Physics, Faculty of Science, Kyoto University, 606-8502, Japan, \\
$^{2}$Institute for Advanced Simulation (IAS-4), Forschungszentrum Jülich, D-52425 Jülich, Germany, \\
$^3$Helmholtz-Institut für Strahlen- und Kernphysik and Bethe Center for Theoretical Physics,
Universität Bonn, D-53115 Bonn, Germany}

\email{jinno.asanosuke.36w@st.kyoto-u.ac.jp}

\recdate{December 22, 2025}

\abst{An analysis of the $\Lambda$ and $\Sigma$ single-particle potentials 
is presented, based on $YN$ interactions derived within chiral
effective field theory up to next-to-next-to-leading order (N$^2$LO). 
The self-consistent Brueckner-Hartree-Fock framework is employed within the continuous choice for the single-particle potential. The result for the 
$\Lambda$ single-particle potential is comparable to the ones obtained with
previous chiral $YN$ interactions up to next-to-leading order (NLO). 
The $\Sigma$ single-particle potential is found weakly attractive, 
in contrast to earlier weakly repulsive results, reflecting new 
constraints from the recent J-PARC E40 data on $\Si^+p$ scattering. 
An estimate of the theoretical uncertainty of the single-particle potentials is provided.}

\kword{$YN$ interaction, hyperons in nuclear matter}

\begin{document}
\maketitle

\section{Introduction}

The hyperon puzzle of neutron stars refers to the problem that 
most of the equations of state that include hyperons are not sufficiently
stiff to support the observed massive neutron stars~\cite{Burgio:2021vgk}. 
Although fifteen years have
passed since the first report of such neutron stars, it is still controversial what mechanisms could prevent
hyperons from appearing in neutron stars, not least due to insufficient constraints on the interactions between
hyperons and nuclear matter. A key ingredient for discussing the appearance of hyperons in neutron stars is the corresponding single-particle potential in 
dense matter, in particular the one of the $\Lambda$ hyperon.

In 2023, a hyperon-nucleon ($YN$) potential based on chiral effective field theory ($\chi$EFT) up to next-to-next-to-leading order (N$^2$LO) was presented \cite{Haidenbauer:2023qhf}, as an extension of previous works
up to next-to-leading order (NLO)~\cite{Haidenbauer:2013oca,Haidenbauer:2019boi}. 
The new interaction builds on a novel regularization scheme, the so-called
semi-local momentum space (SMS) regularization, which has
been shown to work rather well in the nucleon-nucleon ($NN$) sector \cite{Reinert:2017usi}. Furthermore, the potential incorporates new constraints
on the $\Sigma^+ p$ and $\Si^-p$ differential cross sections from
recent J-PARC E40 experiments \cite{J-PARCE40:2022nvq}.

In this contribution, we present selected results from an extended
analysis of the $\Lambda$ and $\Sigma$ 
single-particle potentials $U_\La$ and $U_\Si$ \cite{Jinno:2025vgm}, for the SMS $YN$ interaction up 
to N$^2$LO in combination with $NN$ forces likewise derived in the SMS scheme \cite{Reinert:2017usi}. 
The self-consistent Brueckner-Hartree-Fock method is
employed \cite{Petschauer:2015nea}, within the continuous choice for 
the single-particle potential, and results for the density dependence
of $U_\La$ and $U_\Si$ are reported. 
An estimate of the theoretical uncertainty of the single-particle potentials 
in symmetric nuclear matter (SNM) and in pure neutron matter (PNM) is 
provided. 
 
\section{Results}
\label{sec:Results}

\begin{figure*}[tbhp]
    \centering
    \includegraphics[width=0.89\linewidth]{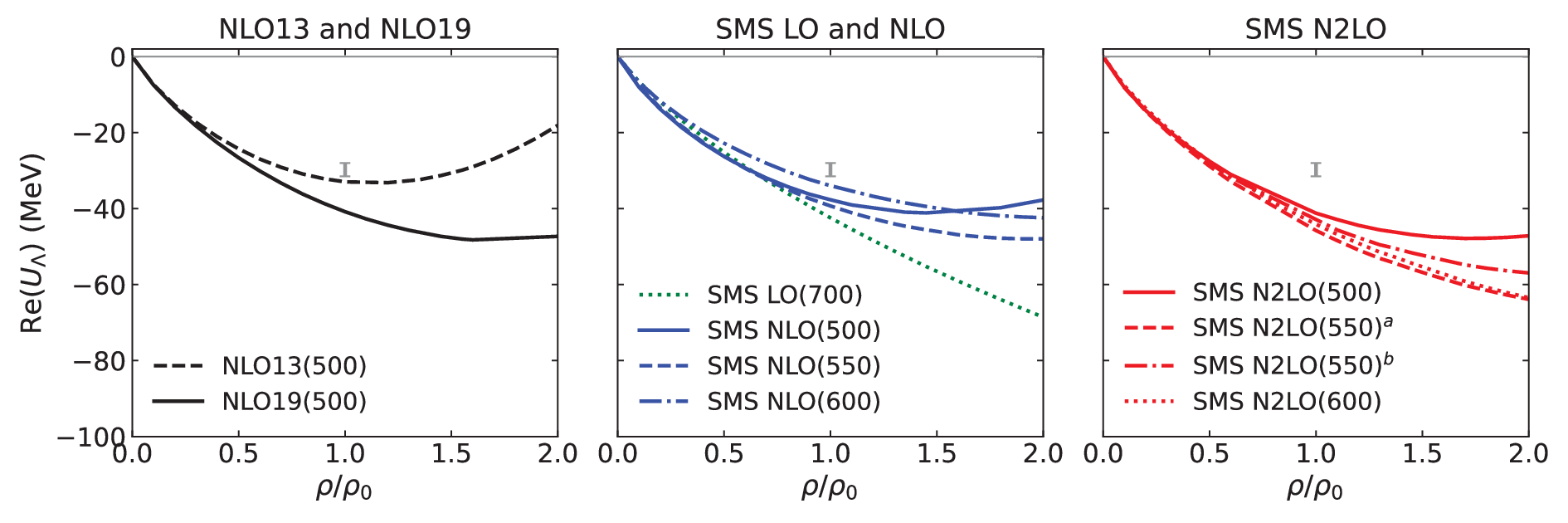}
    \caption{Density dependence of the $\Lambda$ single-particle potential in
    symmetric nuclear matter.
    The bar symbolizes the quasi-empirical value~\cite{Gal:2016boi}. 
    }
    \label{fig:UL_rhoSNM}
\end{figure*}

Results for the $\Lambda$ single-particle potential $U_\Lambda$ in SNM are presented in Table~\ref{tab:La} and in Fig.~\ref{fig:UL_rhoSNM}. 
The notation for the employed $YN$ interactions is adopted from Ref.~\cite{Haidenbauer:2023qhf} and specifies the various potentials by 
their chiral order and by the employed cutoff mass. 
The superscripts $a$ and $b$ in case of SMS N$^2$LO(550) refer to variants
with different $P$-wave interactions~\cite{Haidenbauer:2023qhf}.  
The required nucleon single-particle potential $U_N$ is
evaluated from the SMS N$^4$LO$^+$ $NN$ force~\cite{Reinert:2017usi}
with cutoff $450$~MeV. 

Table~\ref{tab:La} summarizes the results for the $\Lambda$ single-particle potential $U_\Lambda$ 
and its partial-wave decomposition at nuclear matter saturation density 
($\rho_0 = 2k^3_F/(3\pi^2)$, with the Fermi momentum $k_F = 1.35$~fm$^{-1}$), 
and for zero momentum of $\Lambda$. In lowest order of the hole-line expansion,
this quantity represents the binding energy of the $\Lambda$ in infinite nuclear matter
and can be compared with the quasi-empirical value
inferred from hypernuclear experiments, which amounts to $U_\Lambda(\rho_0)\approx-30$~MeV \cite{Gal:2016boi}.
The values of $U_\Lambda(k=0)$ predicted by the 
SMS N$^2$LO potentials range from $-41$ to $-46$~MeV, those of the NLO 
potentials are between $-34$ and $-39$~MeV~\cite{Jinno:2025vgm}.
The predictions for the $YN$ potentials from 2019 (NLO19 \cite{Haidenbauer:2019boi})
and 2013 (NLO13 \cite{Haidenbauer:2013oca}),
$-41$~MeV and $-33$~MeV, respectively, are comparable to the SMS results. 
A recent phenomenological analysis with a 
$\La$-nucleus optical potential~\cite{Friedman:2023ucs} suggests indeed
that the $YN$ two-body contribution to $U_\Lambda(\rho_0)$ is $-38.6 \pm 0.8$~MeV and attributes the difference of $11.3\pm 1.4$~MeV to the 
quasi-empirical 
value to repulsive three-body forces.
In this context, see also the Skyrme model calculation in Ref.~\cite{Jinno:2023xjr} 
which includes likewise many-body effects.

\begin{table*}[tbp]
\renewcommand{\arraystretch}{1.2}
\centering
\caption{
Partial-wave contributions to $ U_\La (k = 0)$ (in MeV)
{at} $k_F = 1.35 \ {\rm fm}^{-1}$. Results are presented for 
the SMS $YN$ interactions at different chiral orders and with 
different cutoffs \cite{Haidenbauer:2023qhf}
and, in addition, for the NLO13(500) \cite{Haidenbauer:2013oca}
and NLO19(500) \cite{Haidenbauer:2019boi}  interactions. 
}
\begin{tabular}{c|rcrrrc|c}
\hline\hline
 & $^1S_0$ & $^3S_1+^3D_1$ & $^3P_0$ & $^1P_1$ & $^3P_1$ & $^3P_2+^3F_2$ &
\, Total \, \\
\hline
SMS LO(700)      & $-13.9$ & $-28.2$ & $-1.0$ & $0.7$ & $1.0$ & $-0.8$ & $-42.4$ \\ 
\hline
SMS NLO(500)     & $-15.7$ & $-25.6$ & $0.4$ & $4.5$ & $2.0$ & $-3.2$ & $-37.7$ \\ 
SMS NLO(550)     & $-15.0$ & $-24.5$ & $0.5$ & $2.5$ & $1.5$ & $-4.1$ & $-39.3$ \\ 
SMS NLO(600)     & $-12.5$ & $-20.2$ & $0.4$ & $1.0$ & $1.5$ & $-4.0$ & $-34.0$ \\ 
\hline
SMS N$^2$LO(500)    & $-16.2$ & $-24.8$ & $0.3$ & $2.9$ & $0.6$ & $-3.2$ & $-41.2$ \\ 
SMS N$^2$LO(550)$^a$  & $-15.2$ & $-26.9$ & $0.1$ & $1.2$ & $0.9$ & $-4.9$ & $-45.8$ \\ 
SMS N$^2$LO(550)$^b$  & $-15.1$ & $-27.3$ & $0.4$ & $1.5$ & $1.6$ & $-3.4$ & $-43.4$ \\ 
SMS N$^2$LO(600)    & $-15.3$ & $-25.4$ & $0.2$ & $1.6$ & $1.0$ & $-5.0$ & $-44.1$ \\ 
\hline
NLO13(500)       & $-15.3$ & $-18.9$ & $0.9$ & $0.2$ & $1.6$ & $-1.3$ & $-33.1$ \\ 
NLO19(500)       & $-13.5$ & $-28.7$ & $0.9$ & $0.3$ & $1.6$ & $-1.2$ & $-40.9$ \\ 
\hline\hline
\end{tabular}
\vskip -0.4cm 
\label{tab:La}
\renewcommand{\arraystretch}{1.0}
\end{table*}

In Fig.~\ref{fig:UL_rhoSNM}, the density dependence of $U_\La$ is presented. 
For the SMS interactions with a cutoff of $500$~MeV, and also for NLO13 and 
NLO19, $U_\La$ reaches a minimum in the considered density range, 
whereas for the SMS NLO and N$^2$LO interactions with larger cutoffs
the minimum occurs at somewhat higher density.
The case of NLO13 is special because there is already a pronounced 
trend towards a repulsive $U_\La$ from around $1.5\,\rho_0$ onwards, while 
a much slower variation with density is observed for the 
other interactions. As known from the work of Gerstung \textit{et al.}, $U_\La$ for 
NLO13 eventually changes sign, i.e. becomes repulsive, 
around $2.5\,\rho_0$~\cite{Gerstung:2020ktv}. 
The origin of the strikingly different behavior of the $U_\La$ results 
for NLO13 and NLO19 has been thoroughly discussed in
Ref.~\cite{Haidenbauer:2019boi}. 
It is a consequence of a marked difference in the strength of the 
$\La N$-$\Si N$ transition potential between the two sets of $YN$ interactions.
In essence, the larger (smaller) channel-coupling strength of NLO13 (NLO19)
leads to less attractive (more attractive) $U_\La$, even when the 
corresponding $YN$ scattering results, including the $\La N$-$\Si N$
transition cross sections, are identical. 
As likewise discussed in Ref.~\cite{Haidenbauer:2019boi}, the strength of 
the $\La N$-$\Si N$ transition potential itself is not an observable and is 
closely connected with possible contributions from $\La NN$ three-body forces.


\vskip -1.0cm
\begin{table*}[htbp]
\renewcommand{\arraystretch}{1.2}
\centering
\caption{
Partial-wave contributions to the real part of
$U_\Sigma(k=0)$ (in MeV) {at} $k_F = 1.35 \ {\rm fm}^{-1}$.
Same description of interactions as in Table \ref{tab:La}.
}
\begin{tabular}{c|rrrrrr|r}
\hline\hline
& \multicolumn{3}{c}{Isospin $I=1/2$} & \multicolumn{3}{c|}{Isospin $I=3/2$} & Total\\
& $^1S_0$ & $^3S_1$+$^3D_1$ & $P$ & $^1S_0$ & $^3S_1$+$^3D_1$ & $P$ \ &  \\
\hline
SMS LO(700)        & 7.1 & $-$16.7 & $-$1.9 & $-$10.4 & 24.7 & $-$1.5 & 1.0 \\
SMS NLO(550)       & 8.0 & $-$25.0 & $-$0.3 & $-$10.7 & 20.3 & $-$3.4 & $-$11.3\\
SMS N$^2$LO(550)$^a$ & 7.5 & $-$24.9 & 3.4 & $-$11.0 & 20.0 & $-$4.4 & $-$10.3 \\
SMS N$^2$LO(550)$^b$ & 7.5 & $-$24.8 & 3.6 & $-$11.0 & 20.0 & $-$5.0 & $-$11.0 \\
\hline\hline
NLO13(500)         & 6.2 & $-$26.2 & 3.7 & $-$11.1 & 32.5 & $-$0.8 & 3.7 \\
NLO19(500)         & 6.0 & $-$19.5 & 3.8 & $-$10.4 & 32.7 & $-$0.8 & 11.2 \\
\hline\hline
\end{tabular}
\label{tab:Sio}
\renewcommand{\arraystretch}{1.0}
\end{table*}

\begin{figure*}[tbhp]
    \centering
    \includegraphics[width=0.31\linewidth]{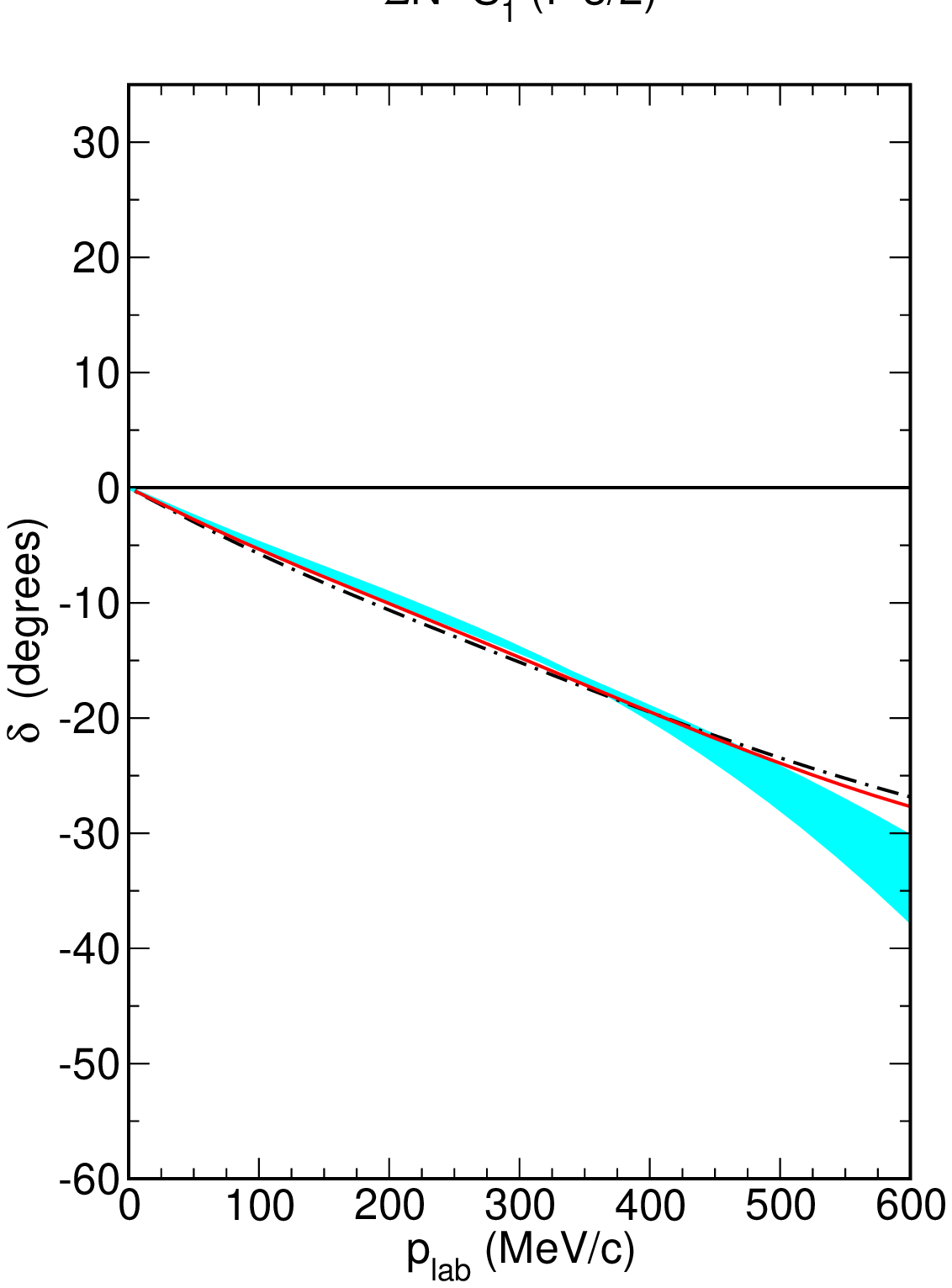}
        \includegraphics[width=0.31\linewidth]{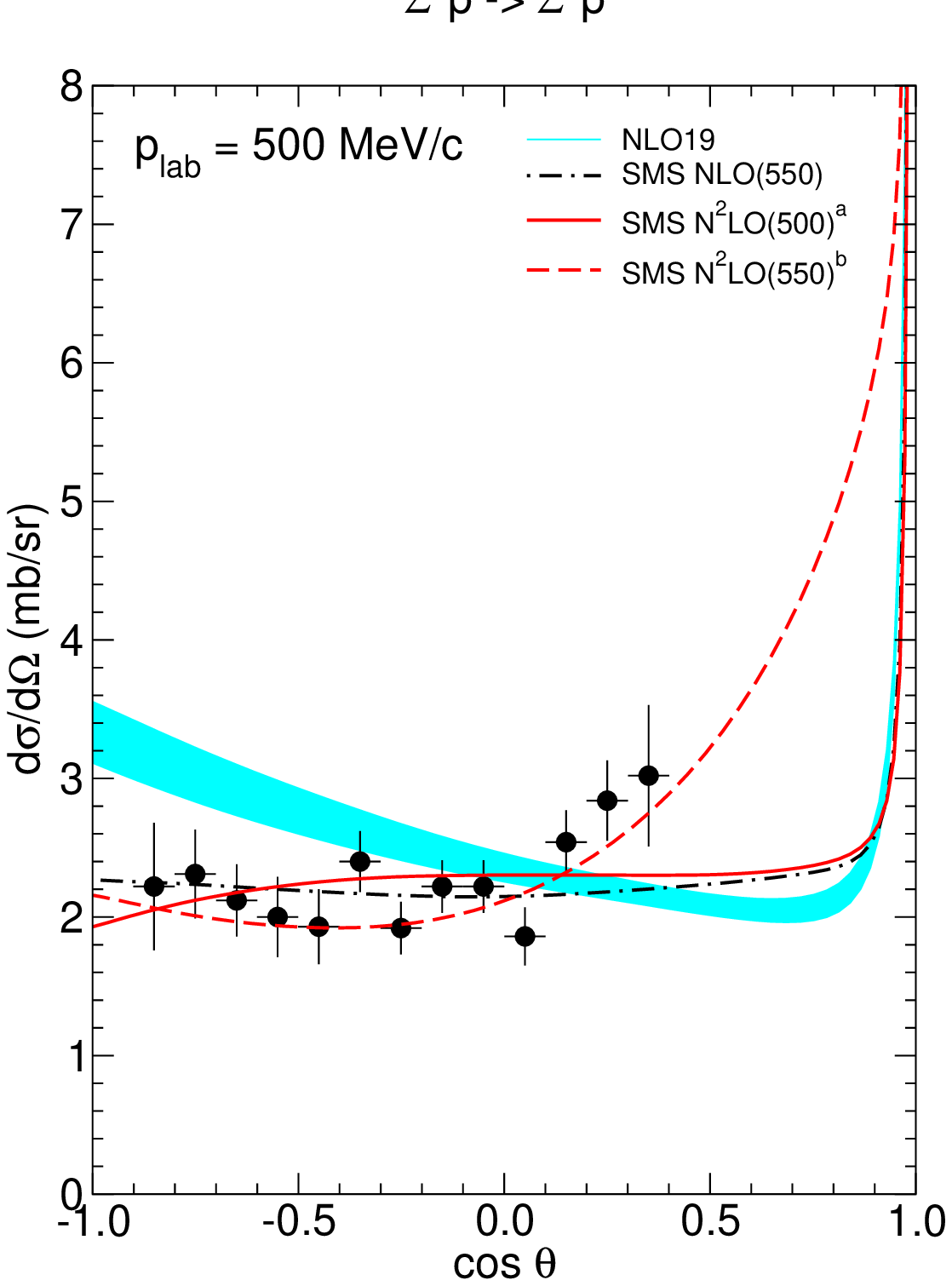}
            \includegraphics[width=0.31\linewidth]{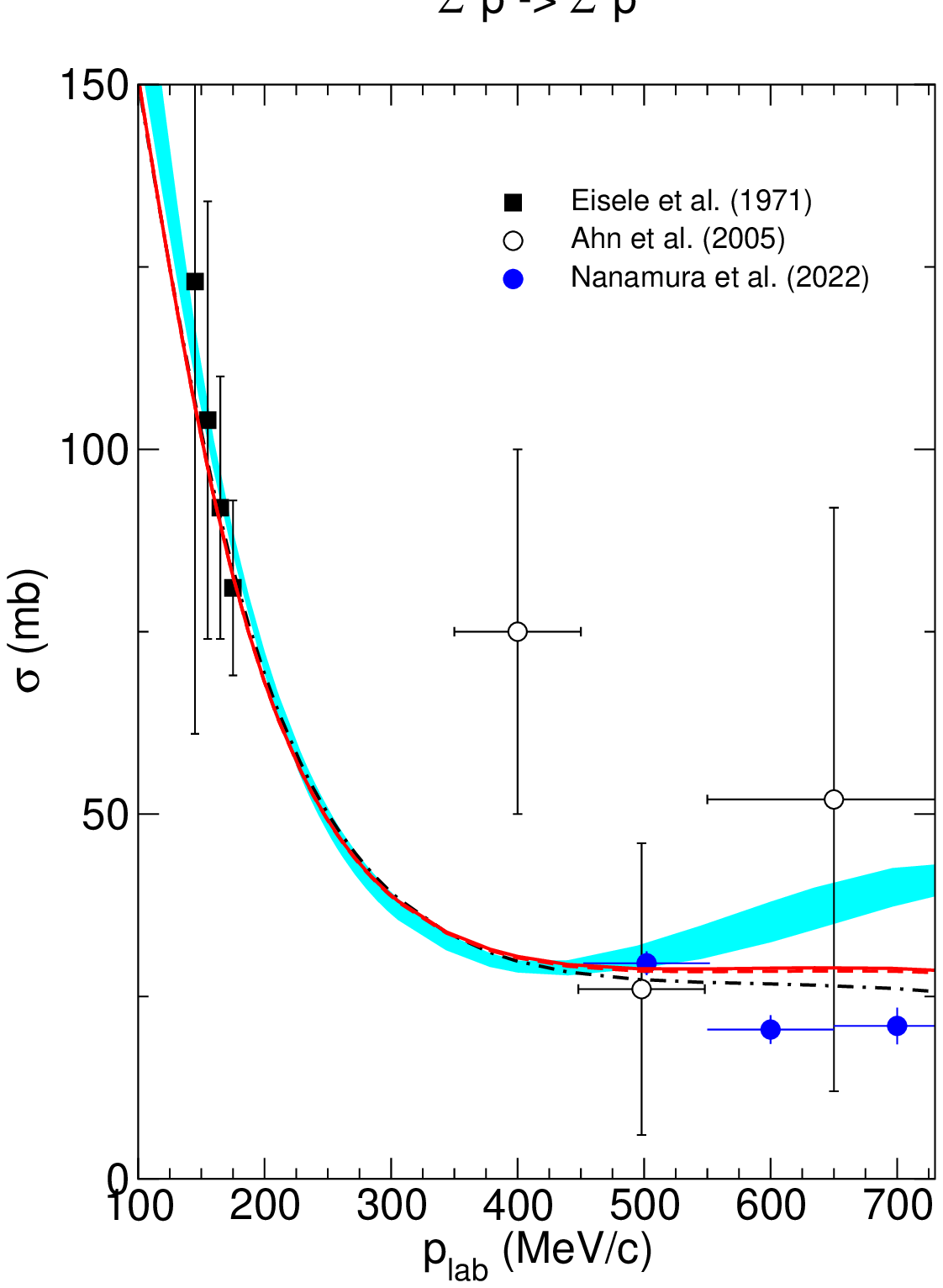}
    \caption{$\Si^+p$ scattering results \cite{Haidenbauer:2023qhf}.  
    Left: $^3S_1$ phase shift. Center: 
    Differential cross section at $500$~MeV/c. Right: Integrated $\Si^+p$
    cross section. Solid, dashed, and dash-dotted lines are the results
    for the SMS N$^2$LO(550)$^a$, N$^2$LO(550)$^b$, and NLO(550) potentials,
    respectively. The band represents the results of the NLO19 potential
    from Ref.~\cite{Haidenbauer:2019boi}. The E40 data~\cite{J-PARCE40:2022nvq}
    are indicated by filled circles. 
    }
    \label{fig:SN}
\end{figure*}

Total results and the partial-wave decomposition for the $\Sigma$
single-particle potential at $\rho_0$ are listed in Table~\ref{tab:Sio}. 
The range for $U_\Si$ deduced from phenomenological analyses of data on $\Si^-$
atoms and 
$(\pi^-,K^+)$
spectra is $10$--$50$~MeV~\cite{Gal:2016boi}.
Obviously, only the SMS LO interaction yields repulsive results, while the 
SMS NLO and N$^2$LO potentials predict an attractive $\Si$ potential.
In contrast, 
NLO13 as well as NLO19 yield a repulsive $U_\Si$ at $\rho_0$. The primary
reason for the difference is that the contribution from the $^3S_1$--$^3D_1$ 
partial wave with isospin $I=3/2$ is noticeably more repulsive for NLO13 and 
NLO19 than that of the SMS $YN$ potentials, as can be seen from the 
isospin decomposition. 
A reduction of the interaction strength in that channel has become 
necessary due to constraints from recent J-PARC E40 data on $\Si^+p$ 
scattering \cite{J-PARCE40:2022nvq}. The $YN$ potentials NLO13 and NLO19
overshoot those data, see Fig.~\ref{fig:SN}.
Specifically, they predict an unrealistic rise in the integrated cross section 
for large momenta (right side) caused by an artificial increase of repulsion
in the $^3S_1$ partial wave (left side).

\begin{figure*}[tbhp]
    \centering
    \includegraphics[width=0.89\linewidth]{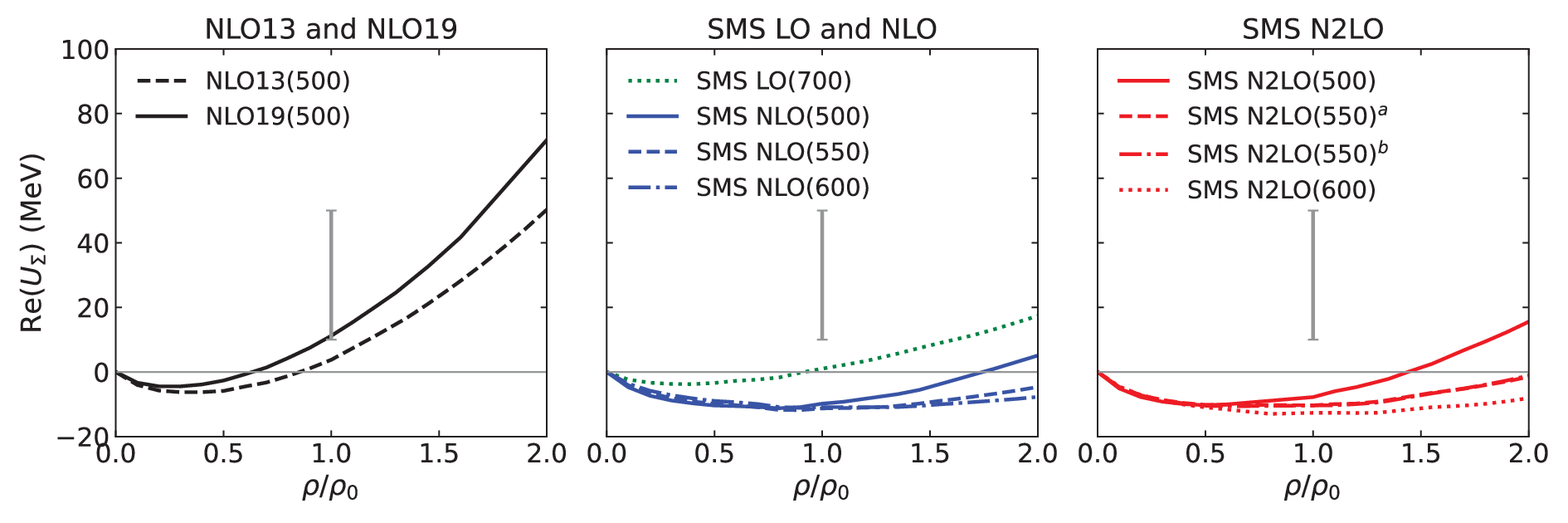}
    \caption{Density dependence of the $\Sigma$ single-particle potential in
    symmetric nuclear matter.
    The bar indicates results from phenomenological 
    analyses \cite{Gal:2016boi}.
    }
    \label{fig:US_rhoSNM}
\end{figure*}

Figure~\ref{fig:US_rhoSNM} shows the density dependence of $U_\Si$. 
The SMS NLO and N$^2$LO potentials predict an attractive $U_\Si$ throughout,
except for the potentials with a cutoff of 500~MeV where $U_\Si$ turns 
to repulsion at around $1.5\,\rho_0$. 
The results for NLO13 and NLO19 are radically different. Here, $U_\Si$ 
becomes repulsive already at lower densities, and moreover the repulsion 
increases rapidly with density. 
As already discussed above, the different behavior 
is a consequence of the constraints by the J-PARC E40 $\Si^+p$ 
data~\cite{J-PARCE40:2022nvq} on the $\Si N$ interaction in the $I=3/2$ 
channel that have been taken into account in the SMS $YN$ potentials. 

\vskip -0.4cm

\vskip -1.1cm 

\section{Uncertainty estimate}
\label{sec:uncertaity}

$\chi$EFT has not only the salient feature that there is an underlying power
counting that allows to improve calculations systematically by going to higher 
orders in the perturbative expansion, it also allows to estimate the 
theoretical uncertainty due to the truncation in that expansion. 
Below, we discuss the truncation error for $U_\La$ and $U_\Si$,
following the method proposed by Epelbaum, Krebs, and Mei\ss ner (EKM) \cite{Epelbaum:2014efa,LENPIC:2015qsz}. 
It combines information about the expected size and the actual size of 
higher-order corrections. 
It can be applied to any observable $X$ that has been evaluated up to
a specific order $i$ in the chiral expansion, $X^{(i)}$.
The concrete expressions for the corresponding uncertainty 
$\delta X^{(i)}$ are~\cite{Epelbaum:2014efa}, 
\begin{eqnarray}
\label{eq:EKM}
	&&\delta X^{\text{LO}} = Q^2 \left|X^{\text{LO}}\right|, \qquad
	\delta X^{\text{NLO}} =\max\Big(
		Q^3 \left|X^{\text{LO}}\right|,
		Q\left|X^{\text{NLO}}-X^{\text{LO}}\right|
	\Big), \nonumber\\
	&&\delta X^{\text{N$^2$LO}} = \max\Big(
		Q^4 \left|X^{\text{LO}}\right|,
		Q^2 \left|X^{\text{NLO}} - X^{\text{LO}}\right|, 
		 Q \left|X^{\text{N$^2$LO}}-X^{\text{NLO}}\right|
	\Big) \ , 
\end{eqnarray}
with the additional constraint for the theoretical uncertainties at LO
and NLO to have at least the size of the actual higher-order
contributions \cite{Epelbaum:2014efa}. 
Here, $Q$ represents the expansion scale in the chiral expansion, 
which is given by $Q\in \{p/\Lambda_b,M_\pi/\Lambda_b\}$,
where $p$ is the typical momentum of the baryons, $M_\pi$ 
is the pion mass, and $\Lambda_b$ is the breakdown scale of the 
chiral expansion.
In a past application of this method to nuclear matter properties by 
Hu \textit{et al.}~\cite{Hu:2016nkw}, the momentum scale $p$ has been identified 
with the Fermi momentum $k_F$. We adopt the same prescription in our work. 
For the breakdown scale, $\Lambda_b = 480$ and $600$~MeV
are used for SNM and PNM, respectively, a choice guided by 
the Bayesian analysis of nuclear matter properties by 
Hu \textit{et al.}~\cite{Hu:2019zwa}.

\begin{figure*}
    \centering
    \includegraphics[width=0.32\linewidth]{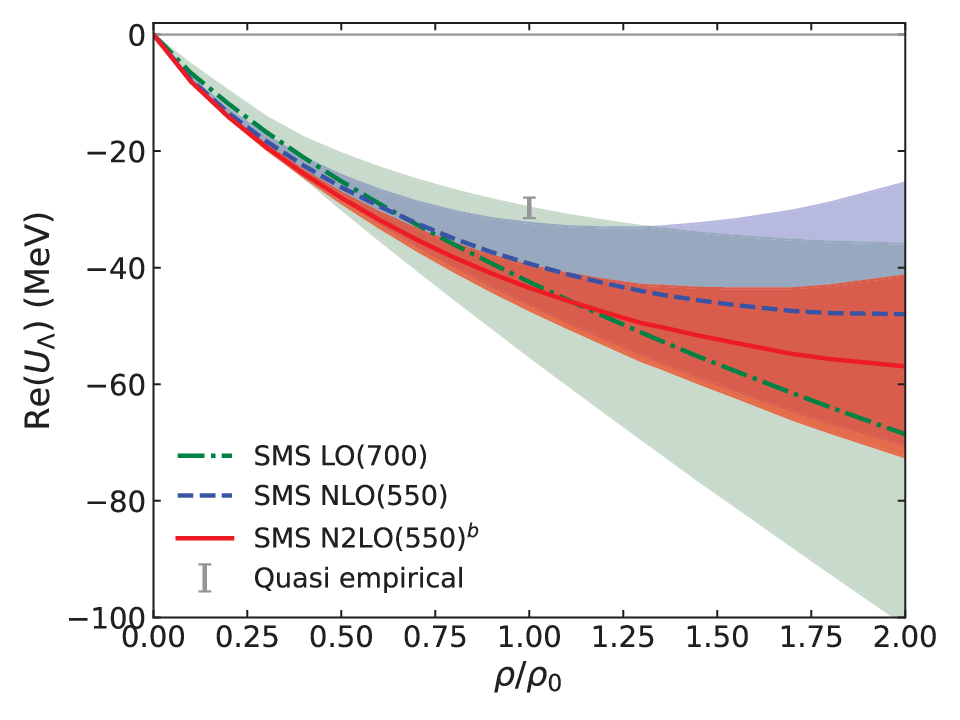}
    \includegraphics[width=0.32\linewidth]{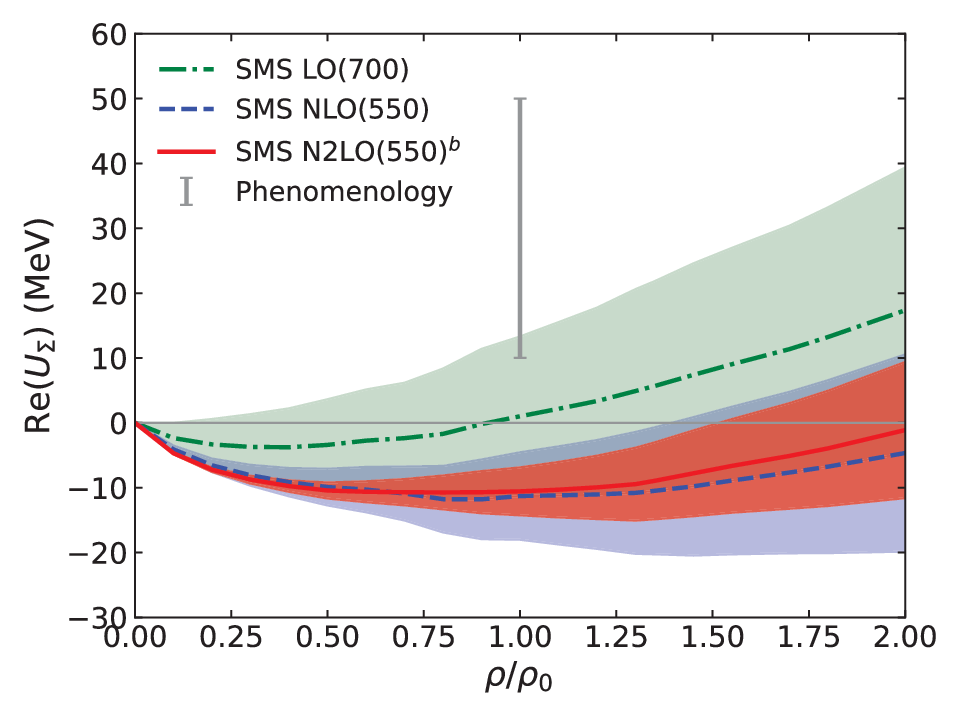}
    \includegraphics[width=0.32\linewidth]{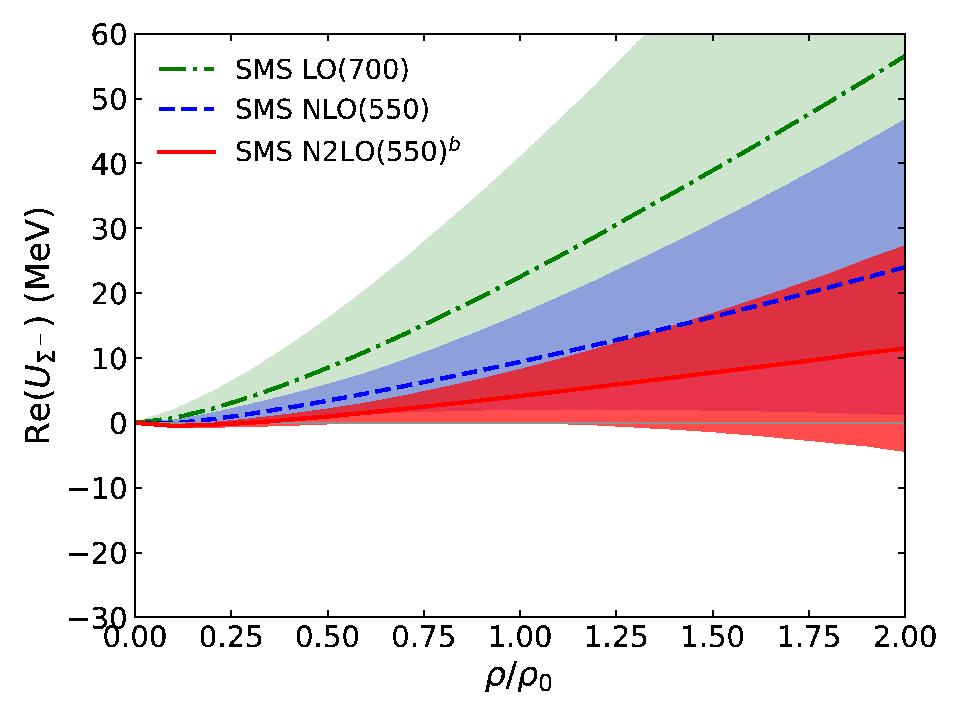}
    \caption{Uncertainty estimate of the $\Lambda$ 
    and $\Sigma$ single-particle potentials in
    symmetric nuclear matter and  of the $\Sigma^-$ single-particle potential in
    pure neutron matter, from left to right. 
    The SMS NLO and N$^2$LO$^b$ interactions with cutoff $550$~MeV are employed, while 
    for SMS LO the potential with $700$~MeV cutoff is used.
    The bars symbolize the values from Ref.~\cite{Gal:2016boi}. 
    }
    \label{fig:uncertainty}
\end{figure*}

In Fig.~\ref{fig:uncertainty}, the uncertainty estimates for the density
dependence of $U_\La (k=0)$ and $U_\Si (k=0)$ are shown.
The SMS LO(700), NLO(550), and N$^2$LO(550)$^b$ $YN$ potentials have
been selected as exemplary set.
The other cutoffs of $500$ and $600$~MeV result in very similar uncertainty bands.
As one can see, the uncertainty of the predicted single-particle potentials increases
rapidly with density and is already substantial at the highest density considered. 

Regarding the $\Lambda$ single-particle potential at $\rho_0$,
the results for the  LO and NLO interactions
are more or less in line with the quasi-empirical value of $U_\Lambda(\rho_0)\approx -30$~MeV 
within the uncertainty estimate, see left side of Fig.~\ref{fig:uncertainty}. 
To be concrete, $U^{\rm NLO}_\Lambda(\rho_0) = -39.3 \pm 7.0~\text{MeV}$. 
The uncertainty of the NLO result provides also a measure for the size of 
possible $YNN$ three-body forces which start to contribute at N$^2$LO.
Interestingly, the present estimate is compatible with the 
actual contribution of an effective three-body force evaluated in
Ref.~\cite{Gerstung:2020ktv}, established within the assumption of 
decuplet saturation and based on LECs constrained from dimensional scaling
arguments. It is also comparable to the value suggested in
Ref.~\cite{Friedman:2023ucs} within a phenomenological analysis. 
The estimate for the N$^2$LO result at $\rho_0$,
$U^{\rm N^2LO}_\Lambda(\rho_0) = -43.5 \pm 3.8~\text{MeV}$,
is not compatible with the quasi-empirical value. Note, however, that
the calculation at N$^2$LO is incomplete, i.e. three-body forces are 
missing! Thus, as a matter of fact, the actual uncertainty is here still 
the one obtained for the result at NLO. 

The uncertainty estimate for $\Sigma$ in SNM is shown in the center 
of Fig.~\ref{fig:uncertainty}. The result at $\rho_0$ is
$U^{\rm NLO}_\Sigma(\rho_0) = -11.3 \pm 6.7~\text{MeV}$ and
$U^{\rm N^2LO}_\Sigma(\rho_0) = -10.6 \pm 3.7~\text{MeV}$.
The uncertainty bands for the NLO and N$^2$LO potentials are similar because 
there is not much difference between the results at NLO and N$^2$LO, as
can be seen from Fig.~\ref{fig:US_rhoSNM}. Obviously, $U_\Si$ is
predicted to be attractive up to $1.5\,\rho_0$, even when considering the
theoretical uncertainty. 
At the right side of Fig.~\ref{fig:uncertainty}, the uncertainty estimate for 
the $\Sigma^-$ single-particle potential in PNM is shown.
The values at $\rho_0$
are $U^{\rm NLO}_{\Sigma^{-}}(\rho_0) = 9.4 \pm 7.2~\text{MeV}$ and
$U^{\rm N^2LO}_{\Sigma^{-}}(\rho_0) = 4.1 \pm 4.0~\text{MeV}$.
Compared to the case of $U_\Sigma$ in SNM, the N$^2$LO uncertainty band
is noticeably larger due to the fact that for a given $\rho_0$ the Fermi 
momentum in PNM is larger. 
Obviously $U_{\Si^-}$ is predominantly repulsive, for the NLO as well
as for the N$^2$LO potential. 

\section{Summary and Outlook}

In this contribution, we reported results for the in-medium properties 
of the hyperons $\La$ and $\Si$, based on a recent $YN$ interaction
up to N$^2$LO in the chiral expansion, established by the J\"ulich-Bonn
group~\cite{Haidenbauer:2023qhf}. 
It turned out that the $\La$ single-particle potentials predicted by the new
NLO and N$^2$LO potentials are slightly more attractive at nuclear matter
saturation density than the quasi-empirical value. 
The $\Si$ single-particle potentials are found weakly attractive at
the saturation density. 

Next, $YNN$ three-body forces have to be included in order to
complete the calculation up to N$^2$LO. Their explicit computation is,
unfortunately, rather challenging since, in principle, one needs to solve the 
Bethe-Faddeev equations. However, as alternative and simplification one can 
consider an approximate treatment by summing one of the nucleons over the
occupied states in the Fermi sea which
then leads to an effective density-dependent two-body interaction 
\cite{Holt:2019bah,Petschauer:2016pbn}. Of course, for being more consistent with
available hypernuclear few-body calculations \cite{Le:2023bfj,Le:2024rkd} the so far 
adopted on-shell approximation \cite{Gerstung:2020ktv,Haidenbauer:2016vfq}
should be avoided.

\end{document}